\documentstyle[preprint,aps]{revtex} 
\begin{document} 

\title{Transmission and
Reflection Studies of Periodic and Random Systems with Gain}

\author{Xunya Jiang and  C.~M.~Soukoulis}

\address{Ames Laboratory and Department of Physics and Astronomy,\\
Iowa State University, Ames, IA 50011\\}

\author{\parbox[t]{5.5in}{\small
 The transmission ($T$) and reflection ($R$) 
coefficients are studied in periodic systems  and random 
systems with gain. For both the periodic electronic tight-binding model 
and the periodic classical many-layered model, we obtain numerically and 
theoretically the dependence of $T$ and $R$.  The critical length of 
periodic system $L_c^0$, 
above which $T$ decreases with size of the system L while $R$ approaches 
a constant value, is obtained to be inversely proportional to the 
imaginary part $\varepsilon''$ of the dielectric function $\varepsilon$.  
For the random system,  $T$ and $R$ also show a nonmonotonic behavior 
versus L. For short systems ($L<L_c$) with 
gain $<lnT>=(l_g^{-1}-\xi_0^{-1})L$. For large systems ($L>>L_c$) with 
gain $<\ln{T}>=-(l_g^{-1}+\xi_0^{-1})L$. $L_c$, $l_g$ and $\xi_0$ are the 
critical, gain and localization lengths, respectively. The dependence of the 
critical length $L_c$ on $\varepsilon''$ 
and disorder strength $W$ are also given. Finally, the 
probability distribution of the reflection $R$ for random systems with 
gain is also examined. Some new very interesting behaviors are observed.   
\\ \\
PACS number(s): 42.25.Bs, 71.55.Jv, 72.15.Rn, 05.40.+j}}

\maketitle
\newpage
\section{Introduction} 

   While the study of localization of classical and quantum waves in 
random disordered media has been well understood 
\cite{Anderson,John,EcoSou,Sou}
, recently, the wave 
propagation in amplifying random media has been pursued intensively 
\cite{experi,Letokhov,PR,Zhang,dual,Zyuzin1,Zyuzin2,tightbind,India}. 
Some interesting results have been predicted, such 
as, the localization length of a  random medium with gain \cite{Zhang}, 
the sharpness of back scattering  
coherent peak  \cite{experi,Zyuzin1}, the dual symmetry of 
absorption and amplification
\cite{dual}, the critical size of the system \cite{Letokhov,Zhang}, 
and the probability distribution of 
reflection \cite{PR}. Numerically, two kinds of models are studied, one is 
the electronic tight binding model \cite{tightbind,India}, the 
other is the many-layered model of classical waves\cite{Zhang}  
. Theoretically, a lot of methods are used to get these results, such as 
the diffusion theory \cite{Letokhov,Zyuzin2} and the  transmission 
matrix method \cite{PR}. 
Most of these studies are for homogeneously random systems  which are 
generated by introducing the disorder into the continuous system, and the 
medium parameter, such as the dielectric constant, is assumed to 
vary in a continuous way \cite{experi,Zhang}. 
But  the periodically correlated random systems which are generated by 
introducing the disorder into a periodic system, such as a
photonic-band-structure , have not been studied adequately.  
\par
  With gain, will such  random systems with periodic background  
behave similar as the homogeneously random system? Both experimentally and 
theoretically, the study of such system is very important in understanding 
the propagation of light in random media.  These type of  
photonic-band-structure 
systems are widely used in experiments \cite{EcoSou,Yariv}. 
Theoretically, just as S.John \cite{EcoSou} 
argued, the localization of a photon is from a subtle interplay  
between order and disorder. For the periodically correlated 
random systems with gain, the periodic background plays the order role, 
and  now its interplay  with not only disorder but also with gain should 
be a very interesting new topic.      
\par
  In this paper we address both the electronic tight binding model and 
the many-layered model of classical waves. We first compare the numerical 
results of periodic amplifying system with what 
we can predict theoretically by the transfer matrix method. It's 
surprising to get most of the universal 
properties, such as critical length and exponential decay of 
transmission, of 
homogeneously random system \cite{Zhang} from a 
periodic system too.

 With  the help of some theoretical arguments and numerical results, 
we suggest that the length ${\xi_1}=|1/Im(K)|$, where $K$ is the Bloch 
vector in a periodic system with gain, to replace the gain length 
${l_g}=|1/Im(k)|$  introduced in \cite{Zhang}. This is more 
reasonable 
since the correlated scatterers in a periodic system can make 
the  paths of wave propagation much longer in the system. We also 
think that it is actually the Bloch wave instead of the plane wave which 
propagates in the system. 
Then we introduce disorder into these periodic systems and calculate 
their properties. Our numerical simulations for 
both 
models show that  periodically correlated random systems give  similar 
behaviors
as that of the homogeneously random systems studied previously.
 But in some cases, we get 
 new 
 interesting results for the localization length $\xi$, the critical 
length $L_c$  and the probability distribution $P(R)$ of reflection. All 
these new results are related to the periodic background of such systems. 
We also examine the results of the transmission coefficient $T$ for $short$ 
 $(L<{L_c})$ systems. Our 
numerical results show that the formula of the transmission coefficient of 
media with 
$absorption$ can be generalized to the transmission coefficient of short 
systems 
with $gain$, if we replace the gain length $l_g$ (or $\xi_1$) with 
the negative of the absorption length $-l_a$ in the formula.
To explain our new results of the critical length $L_c$, we compare the two 
basic theories for obtaining the critical length, the Letokhov theory 
\cite{Letokhov} 
and the Lamb theory \cite{LaserPhys},  and we get some new 
theoretical results of critical length which are in good agreement with our 
numerical results. 
The behavior of the distribution of the probability of reflection $P(R)$ is 
much 
more complex than the theoretical prediction of homogeneously random 
system \cite{PR}. We find that the  periodic background influences strongly 
the general behavior of $P(R)$.

\par

The paper is organized as follows. In Sec.II we introduce the two 
theoretical models we are studying. The results for the periodic systems 
with gain are presented in Sec.III, while in Sec.IV the results for the 
random systems with gain are given. Also in Sec.IV we present our new 
theoretical and numerical results for the critical length $L_c$. In Sec.V 
the results for the probability distribution of reflection coefficient 
$R$ for both models are presented. Finally, Sec.VI is devoted to a 
discussion of our results and give some conclusions.

\section{Theoretical Models}  
\subsection{ Many-layered model of classical wave}

Our periodic many-layered model of classical wave consists of two types of 
layers with 
dielectric constant ${\varepsilon_1}={\varepsilon_0}-i\varepsilon''$  and 
${\varepsilon_2}=\chi{\varepsilon_0}-i\varepsilon''$ and thicknesses 
$a=95nm$ and ${b_0}=120nm$ respectively, where the negative part of 
dielectric constant, i.e. $\varepsilon''>0$, denotes the homogeneous 
amplification of the field. We have tried a lot values for $\chi$, such 
as 1.5, 2,  3, 5, 6,   and 
get no essential difference in our results for different values. In this 
paper we choose 
$\chi=2$, i.e. $Re(\varepsilon_1)=\varepsilon_0=1$ and 
$Re(\varepsilon_2)=2\varepsilon_0=2$. The  system has L cells. Each cell is 
composed by two layers with dielectric constant  ${\varepsilon_1}$ and 
${\varepsilon_2}$ respectively.   
Without gain,  we obtain that the wavelength range of the second band of 
this 
periodic system is from 247 nm to 482.6 nm ( the first band has a range from 
592 nm to infinite). So we 
choose the wavelength 360 nm to represent band center, 
the wavelength 420nm as a general case, and the wavelength 470 nm to 
represent the band edge. 

To introduce disorder, we 
choose the width of second layer of the $n$th cell to be random variable 
 ${b}_n={b}_0(1+W\gamma)$, where $W$ describes the 
strength of randomness and $\gamma$ is a random number between $(-0.5, 
0.5)$.  
The whole system is embedded in a homogeneous infinite  material with 
dielectric constant equal to ${\varepsilon_0}$.

For the 1D case, the time-independent Maxwell equation can be written as:
\begin{eqnarray}
     \frac{{\partial^2} E(z)}{\partial {z^2}} + 
{\omega^2\over{c^2}}\varepsilon(z)E(z)=0        \label{Maxw}
\end{eqnarray}

Suppose that in the medium with dielectric constant $\varepsilon_1$  and the 
medium 
with dielectric constant $\varepsilon_2$, the electric field \cite{Zhang} is 
given by the following expressions: 
\begin{eqnarray}
E_{1n}(z) &=& {A_n}{e^{ik(z-{z_n})}}+{B_n}{e^{-ik(z-{z_n})}} \nonumber \\ 
E_{2n}(z) &=& {C_n}{e^{ik(z-{z_n})}}+{D_n}{e^{-ik(z-{z_n})}} \label{feild} 
\end{eqnarray}

Using the appropriate  boundary condition (continuity of the electric 
field $E$ and of the derivative of $E$ at the interface), we obtain that:

\begin{eqnarray}
\left( \begin{array}{c} {A_{n-1}} \\ {B_{n-1}} \end{array} \right) =
\left( {M_n} \right)
\left( \begin{array}{c} {A_{n}} \\ {B_{n}} \end{array} \right) 
\end{eqnarray}
where
\begin{eqnarray}
\left( {M_n} \right) =  
\left( \begin{array}{clcr} 
{e^{-ika}}(cos(q{b_n})-{i\over2}
({\frac{k}{q}}+{\frac{q}{k}})sin(q{b_n}) &
-{i\over2}{e^{-ika}}({\frac{k}{q}}-{\frac{q}{k}})sin(q{b_n}) \\
{i\over2}{e^{ika}}({\frac{k}{q}}-{\frac{q}{k}})sin(q{b_n}) &
{e^{ika}}(cos(q{b_n})+{i\over2}
({\frac{k}{q}}+{\frac{q}{k}})sin(q{b_n})
\end{array} \right)             \label{tranM}
\end{eqnarray}
where $ k={\frac{\omega}{c}}\sqrt{\varepsilon_1} $ and 
$q={\frac{\omega}{c}} \sqrt{\varepsilon_2} $.

From the product of these matrices, $M(L)={\prod_{1}^L}{M_n} $, we can 
obtain the transmission and reflection amplitudes of the sample, 
$t(L)=\frac{1}{M_{11}}$ and $r(L)=\frac{M_{21}}{M_{11}}$.  
For each set of parameters $(L, W, {\varepsilon''})$, the reflection 
coefficient $R=|r|^2$ and the transmission coefficient $T=|t|^2$ 
 are obtained from a large number of random configurations. We have used 
10,000 configurations to calculate the different average values of 
$R$ and $T$, and 1,000,000 configurations to obtain 
$P(R)$. 
Our numerical results 
show  that the localization length  for a system without gain behaves  
$\xi_0 \propto 
1/{W^2}$ for this model, and are in agreement with previous workers.

\subsection{ Electronic tight-binding model}
\par

For the electronic tight-binding model, the wave equation can be written as:
\begin{eqnarray}
\left( \begin{array}{c} {\phi_{n+1}} \\ {\phi_n} \end{array} \right)
=\left( {M_n} \right) \left( \begin{array}{c} {\phi_n} \\ {\phi_{n-1}} 
\end{array} \right) 
\end{eqnarray}
where
\begin{eqnarray}\label{tim}
\left( {M_n} \right)=
\left( \begin{array}{clcr} E-{\epsilon_n} & -1 \\
                           1       &  0  \end{array} \right)
\end{eqnarray}
${\epsilon_n}=W{\gamma}-i\eta$, where W describes the strength of 
randomness, $\gamma$ is a random number between $(-0.5, 0.5)$ ,
$\eta>0$  corresponds to amplification and $\phi_n$ is  
the wave function at site $n$. The length L of the system is 
the total lattice number of the system. The system is embedded 
in two identical semi-infinite perfect leads on either side. For the left 
and the right sides, we have ${\phi_0}=1+r(L)$ and 
${\phi_{L+1}}=t(L){e^{ik(L+1)}}$.  We can obtain reflection amplitude $r(L)$ 
and transmission amplitude $t(L)$ by the products of matrices,  
$M(L)={\prod_{1}^L}{M_n} $. \begin{eqnarray}
t(L)=\frac{-2isin(k)}{{M_{11}}{e^{-ik}}+{M_{12}}-{M_{22}}{e^{ik}}-{M_{21}}}
{e^{-ik(L+1)}} \nonumber \\
r(L)=\frac{{M_{21}}{e^{ik}}+{M_{22}}-{M_{11}}-{M_{12}}{e^{-ik}}}
          {{M_{11}}{e^{-ik}}+{M_{12}}-{M_{21}}-{M_{22}}{e^{ik}}}
          {e^{-ik}} \label{tTR}
\end{eqnarray}
where $ k=arccos(\frac{E}{2})$.  

When $W=0$ and without gain, the model is a periodic one with only one 
band spanning in energy between -2 and 2. Notice that the hopping matrix 
elements in Eq.(\ref{tim}) are equal to one, which is our unit of energy. 
So we choose 
$E=0$ to represent band center, $E=1$ as a  general case, $E=1.8$ to 
represent band edge.   

Similar as the many-layered 
model, for each set of parameters $(L, W, \eta)$, 10,000 
random configurations were used to obtain  a 
average value of  $R$ and $T$, and one million random configurations for 
$P(R)$.
Theoretical and numerical results give  that the localization 
length for a system without gain behaves $\xi_0 \propto 1/{W^2}$, in 
agreement with previous workers.

\section{Periodic Systems}

Almost all the properties of the periodic systems of both the many-layered 
model and the tight-binding model can be predicted theoretically.

 \subsection{Classical many-layered model}
\par

For long systems ($L\gg {L_c^0}$) of the many-layered model we have that:
 \begin{eqnarray}
     \lim_{L \rightarrow \infty}  \frac{\partial \ln{T}}{\partial
 L}&=&
-{{\xi_1}^{-1}} \nonumber \\
                                    &=&2Im(K) \propto \varepsilon''  
\end{eqnarray}
where $K$ is the Bloch vector which is a complex number now, and satisfies 
$cosK=cos(ka)cos(q{b_0})
-{\frac{1}{2}}(\frac{k}{q}+\frac{q}{k})sin(ka)sin(q{b_0})$.
Because $ Im(K)<0$, the transmission coefficient $T$ is decaying  
exponentially  for a long system.

For a short system, $ L < {L_c^0} $, we have:
\begin{eqnarray}\label{pct}
\frac{\partial \ln{T}}{\partial L} =
1/{\xi_1'} & \simeq &  2|C||Im(K)| \nonumber \\
                             & \simeq & 2|Im(K)|
\end{eqnarray}  
Where $C=-[sin(ka)cos(q{b_0})+cos(ka)sin(q{b_0})]/[2sin(K)] $, and $|C|$ is
 larger than but very close to 1 when wavelength is at the band center,
and become bigger when wavelength approached the band edge.

So the slope of $\ln{T}$ vs $L$ for a short periodic system is almost same
as the negative value of the slope for the long system.   The slopes of 
$\ln{T}$ at  both sides of the maximum
are approximately symmetric. 
In Fig.~1a, we can see that, when $L < L_c^0$, $T$  increases vs L with 
the slope  $1/\xi_1'$, and get to a maximum at $ L_c^0$ and  decays 
exponentially when $L >L_c^0$ with the slope $1/\xi_1$.

From the  behavior of the theoretical expressions of $T$ or $R$ , when 
$T$ or $R$ goes to infinite,  we can obtain analytically that $L_c^0$ is 
given as: 

\begin{eqnarray}\label{LC0}
{L_c^0} \simeq {\xi_1}\ln {\left(\frac{|C|+1}{|C|-1}\right)} 
\end{eqnarray}
where $C$ is same as defined above in Eq.(\ref{pct}), and $|C|$ is 
close to one. From the property of $|C|$ discussed above, we can see
that
$ {L_c^0} >> {\xi_1}$ at the band center , and becomes smaller when 
wavelength approaches 
the band edge. The  value of $|C|$ is almost independent of gain, so 
${L_c^0}$
is parallel to $1/\varepsilon''$ or $\xi_1$. We have shown that 
$L_c^0\varepsilon''$ is almost a constant for a given wavelength, and 
our numerical results agree very well with the theoretical prediction.

The reflection coefficient gets to a maximum value at $L_c^0$ too, and 
fluctuates a 
lot with the size $L$ of the system. When $L$ approaches infinity, $R$ 
reaches a saturated value.  The saturated value of $R$ is given by: 
\begin{eqnarray}
\lim_{L \rightarrow \infty} R =R_0 \simeq 
\frac{{|(\frac{k}{q}-\frac{q}{k})sin(qb)|}^2}{{|sin(K)(|C|-1)|}^2}
\end{eqnarray}
So $R_0$ is almost independent of gain or $\xi_1$.  Fig.~2 shows that 
indeed $R$ 
increase when $L<L_c^0$, gets to its maximum value  and fluctuate 
violently at 
$L_c^0$, then approaches a saturated value which is almost independent of gain.

\subsection{ Electronic tight-binding model}

\par
For the electronic tight-binding model, when $E=0$, the $\ln{T}$ for the long 
system can be obtained by the use of Eq.(7):

\begin{eqnarray}
 \lim_{L \rightarrow \infty} \frac{\partial \ln{T}}{\partial L} 
=-{1/\xi_1}  \simeq  -\eta 
\end{eqnarray}

\par
Similarly as the classical many-layered model, for short system of 
tight-binding model we have:
\begin{eqnarray}
\frac{\partial \ln{T}}{\partial L}=1/{\xi_1'} \simeq \eta
\end{eqnarray}
So the slope symmetry of $\ln{T}$ at both sides of $L_c^0$ still exists.
In Fig.~1b, we can see a similar behavior as in Fig.~1a, when $L<L_c^0$, 
$\ln{T}$ change vs $L$ with the slope of $1/\xi_1'$ and gets maximum at 
$L_c^0$, 
then it begin to decay exponentially with a slope of $1/\xi_1$.

Assuming  that the theoretical expression of $T$, given by Eq.(7), is 
infinite, we can obtain that  $L_c^0$ is given by: 
\begin{eqnarray}
{L_c^0}  \simeq {\frac{2}{\eta}}(\ln{4}-\ln{\eta}) \simeq 2{\xi_1}\ln{(4{\xi_1})} 
\end{eqnarray}
We also shown that ${L_c^0}\eta+2\ln{\eta}$ vs $\eta$, for $E=0$, is a 
constant for different gain and indeed 
find out that the theoretical prediction given by Eq.(14) agree very well 
with the numerical results. 

The reflection coefficient $R$ approaches a saturated value as L goes
to infinite, but the
saturated value of $R_0$ is not a constant independent of the gain as in the
case of
classical many-layered model.  This is clearly seen in  Fig.~3 where we
plot $\ln{R}$ vs L. Notice that the
 $\ln{R}$ curves increase vs $L$ when $L<L_c^0$,  get to a maximum at
$L_c^0$, and then approach a saturated value when L goes to infinity.
Similar results were obtained for $E \neq 0$.

\section{Random Systems}

In Fig.~4, we give the general behavior of 
average value $<\ln{T}>$ vs $L$ for both models. 
We can see the different behaviors for $L<L_c$ and $L>L_c$. When 
$1/\xi_1> 1/\xi_0$ and $L<L_c$, 
$<\ln{T}>$ increase vs L from origin with a slope  which is defined as 
$1/\xi'$ and when 
$L>L_c$, $<\ln{T}>$ decrease vs L with a slope $-1/\xi$. But when $1/\xi_1< 
1/\xi_0$, $<\ln{T}>$ will decrease monotonically, at first with the slope 
$1/\xi_1=-1/|\xi_1|$, at $L_c$, there are a turning point and slope 
changes to $-1/\xi$.  We 
will study the values of $\xi'$ , $\xi$ and $L_c$ in this section.

It was first suggested by Zhang \cite{Zhang} that the localization length 
$\xi$ of a long random system with gain will become smaller than the 
localization length $\xi_0$ of the random system without gain. In 
particular he suggested that:  
      
\begin{eqnarray}
\frac{1}{\xi}=\lim_{L \rightarrow \infty}  
\frac{\partial \ln{T}}{\partial L}=\frac{1}{\xi_0}+\frac{1}{\xi_1} 
\label{locleng} 
\end{eqnarray}
where ${\xi_0}$ is the localization length of the system without gain,  
${l_g}$ is  replaced by ${\xi_1}$ in the original formula of Zhang 
because of the periodic background of our systems. 
  
We have numerically calculated $1/\xi$ for different cases of 
disorder, gain and frequency(energy), and compare it with 
$1/{\xi_0}+1/{\xi_1}$, as shown in Fig.~5a and 5b for 
 the many-layers model and the tight-binding model respectively. For most 
of the cases, Eq.(\ref{locleng}) is a very good formula. Only when 
wavelength is on band edge $and$ when both gain and randomness are very 
strong, we can see that the numerical results deviate  
from the theoretical prediction, which is the solid line in both Fig.~5a 
and~5b.

\par

For short system($L<L_c$), the behavior of $<\ln{T}>$ vs $L$ is quite 
different from that of long system as shown in Fig.~4.  Freilikher 
et.al. and Rammal and Doucot \cite{short} 
obtained that the transmission coefficient of a random system with 
$absorption$ is given by:

\begin{eqnarray} 
<\ln{T}>=(-\frac{1}{l_a}-\frac{1}{\xi_0})L \label{absorb0}
\end{eqnarray}

where  ${l_a}$ is the absorption length and ${\xi_0}$ is the localization 
length. For a medium 
with $gain$, can we just substitute the $-{l_a}^{-1}$ with  ${l_g}^{-1}$ 
 in the Eq.(\ref{absorb0}) to get 
following equation? 

\begin{eqnarray}
\frac{<\ln{T}>}{L}=\frac{1}{\xi'}=(\frac{1}{l_g}-\frac{1}{\xi_0}) \label{ab}
\end{eqnarray}
Because of the periodic background of our models, we use $\xi_1$ to replace 
$l_g$ in our calculations.

So far there is no independent verification for this conclusion. After 
substituting $\xi_1$ for $l_g$, our numerical results show 
that Eq.(\ref{ab}) is correct for $short$ systems with $gain$ for both 
models.  
When the strength of the disorder is a constant , so $\xi_0$ is a constant,
according to Eq.(\ref{ab}),  $1/{\xi_1}-1/{\xi'}$ should 
be equal to $1/{\xi_0}$ and be  constant as the gain varies. 
We have checked this prediction and find indeed that the numerical values are 
almost the same as the ones predicted theoretically. 

From Eq.(\ref{ab}), we can predict the basic features of the 
length dependence of $<\ln{T}>$ shown in Fig.~4. When
$1/{\xi_1} > 1/{\xi_0}$, 
$<\ln{T}>$ will increase with $L$, and will reach a maximum value when $L$ 
gets to $L_c$. But if 
$1/{\xi_1} < 1/{\xi_0}$, the $<\ln{T}>$ will decrease monotonically, 
at first with a slope of $-|1/{\xi_1}-1/{\xi_0}|$
 from the origin, 
at $ {L_c} $ the curve has a turning point and  the slope changes to 
$-|1/{\xi_1}+1/{\xi_0}|$. If $1/{\xi_1} \simeq 1/{\xi_0}$, 
the curve is almost horizontal for small L and begins to decrease with a 
slope 
$-1/\xi$ at the critical length. This behavior is exactly shown in Fig.~4.

\par
The critical length $L_c$ is  one of the most important parameters of a 
random system with gain. 
For a random system, one of the most important theories is the Letokhov
theory \cite{Letokhov}.  Zhang \cite{Zhang}
generalized the theory and used the no-gain localization length $\xi_0$ 
to replace the
diffusion coefficient $D$ in the Letokhov theory and obtain that the 
critical 
length $L_c \simeq \sqrt{ {\xi_0} {l_g} }$, so that we can clearly 
see localization effects in the system. But as shown above, there is a
finite critical length in periodic system when the no-gain localization 
length $ \xi_0$ goes to infinite, so there must be 
other mechanisms for determining the critical length in those systems. We 
find that when the localization effect is strong
enough so that the no-gain localization length $ \xi_0 \ll 
{(L_c^0)}^2/\xi_1$, then the results of the Letokhov theory are quite good. 
But 
when the system randomness is  weak so that $ \xi_0 $ is larger than
$(L_c^0)^2/{\xi_1}$, then the Letokhov theory results are not correct, and 
we have to use other theories, such as 
the Lamb theory \cite{LaserPhys}, which is well known is laser physics, to 
determine $L_c$. Next we will compare the Letokhov theory 
with the Lamb theory, and find the expressions of the critical length in 
different cases.

\par

According to the Letokhov theory \cite{Letokhov,Zyuzin2} the field in 
the system satisfies:

\begin{eqnarray}
\frac{\partial \phi({\vec r}, t) }{\partial t} =
D{\nabla ^2} {\phi({\vec r}, t) } 
+\frac {c{\phi ({\vec r}, t) }}{l_g}             \label{Letokhov} 
\end{eqnarray}
where $D$ is the diffusion coefficient, $c$ is the speed of the wave.

Considering the relaxation  after long time, the solution \cite {Zyuzin2} of 
Eq.(\ref{Letokhov}) is : 

\begin{eqnarray}
\phi({\vec r}, t) 
&\propto& e^{-t[D{(\frac{\pi}{L})}^2 - \frac{c}{l_g}]} \nonumber \\
\frac {\partial \phi({\vec r}, t) } {\partial t} &=&
-D{(\frac{\pi}{L})}^2 \phi({\vec r}, t) 
+\frac {c{\phi ({\vec r}, t) }}{l_g}
\end{eqnarray}

When $L={L_c}=\pi \sqrt{ \frac{D {l_g}}{c}} $ , the system is at a 
critical point. If $L<L_c$ the field will decay vs time, but if $L>L_c$ 
then the field in the system will become stronger and stronger with time. 

We can clearly see that  the physical 
meaning of $L_c$ is the balance point of the gain and loss in the system.
When the $L$ is less than $L_c$, the photon escaping rate, which is 
determined by $\frac{D {\pi^2}}{L^2}$, is 
larger than the photon generating rate, which is determined by 
$\frac{c}{l_g}$ of the system, so the 
photons generated by the stimulated emission can escape from the system 
instantaneously and the system can get to the static state after 
a long time. If $L$ is larger than $L_c$, gain 
is larger than loss,  and
photons will be accumulated in the system \cite{Zyuzin2}. 
Based on the Letokhov theory and the weak localization theory 
\cite{Anderson,John} results, Zhang \cite{Zhang} 
generalized the $L_c$ to be $ L_c \simeq \sqrt{ {\xi_0} {l_g} } $ since 
$D = \frac{1}{3}lc$, where $l$ is the mean free path and 
$\xi_0 = (2\sim4)l$.

\par
In our models, considering the periodic background, we substitute $\xi_1$  
for $l_g$ first. But when the disorder becomes weaker and weaker,
the system  become almost periodic, $\xi_0$ goes to infinite,  
 $L_c$ goes to  $L_c^0$ instead to infinite. How one can explain this 
behavior of $L_c$?  The Lamb theory can give a 
theoretical explanation of it. 
In the Lamb theory, a phenomenological parameter Q(L), 
the quality factor  which generally is a function of system length 
$L$, is introduced to show the energy loss 
rate of the system (also can be thought as the photons loss rate of the 
system). In the Lamb theory, the magnitude of the electric field in a 
linear medium satisfies the following equation:

\begin{eqnarray} \label{Lamb}   
\frac {\partial |E(t)|} {\partial t} = -{\frac {\omega} {2Q(L)}} |E(t)|  
+\frac{c}{l_g} |E(t)|   
\end{eqnarray}

At the critical condition, the gain term and loss term are equal. We have:
\begin{eqnarray}\label{LambCritical}
\frac {\omega} {2Q({L_c})}=\frac{c}{l_g}
\end{eqnarray}

If we compare this Eq.(\ref{Lamb}) with the solution of Letokhov theory 
Eq.(19), we can 
find the similarity between them. This similarity is from the same 
physical principle, the interplay of loss and gain in the system. From 
Eq.(\ref{Lamb}) we can see that the gain term is same as the one given by the
Letokhov theory, the only 
difference is from loss term. Generally, ${\frac {\omega} {2Q}}$ is a 
function of the system length, e.g. for Fabry-Perot interferometer 
${\frac {\omega} {2Q}}\propto {\frac{1}{L}}$ \cite{LaserPhys}. For periodic 
system $Q={Q_p}$, we 
have ${\frac {\omega} {2Q_p}}\propto {\frac{1}{L}}$ too. From the balance 
of gain and loss,  we can get that ${L_c^0} \propto {\xi_1}$ in 
agreement with our results presented in 
section III. This means that in a periodic system the rate of loss is 
not infinite, although the no-gain localization length goes to infinite. The 
rate is 
determined by the ${Q_p}$ of the system and we can get a finite $L_c^0$ 
correspondingly. From the $L_c^0$ obtained above and the critical condition 
given by Eq.(\ref{LambCritical}), we have that the quality factor of the 
periodic system is given by:

\begin{eqnarray} \label{Qp} 
{Q_p}={\frac {\omega {\xi_1}} {2c{L_c^0}} } L
\end{eqnarray}
which is independent of no-gain localization length $\xi_0$.

For a random system, things are a little more difficult.  
The theory of Letokhov doesn't give the detailed information of 
localized modes but it gives a localization related quantity $D$, the 
diffusion coefficient. 
According to the localization theory, $D$ is directly related with 
localization 
length $\xi_0$, just as Zhang discussed \cite{Zhang}. Based on the 
correct results of the   
Letokhov theory in the strong localization case, we can assume that when 
the disorder is strong enough ${\xi_0} 
\ll {{L_c^0}^2}/ {\xi_1} $, the localization effects will dominate 
the escape rate of photons of the system (Our numerical results shown in 
Fig.~11 support this   
assumption).  Lamb theory gives  
 that the Q of a strong random system is determined by the 
localization effect. By comparing the corresponding  terms in Eq.(19) 
and Eq.(\ref{Lamb}), we obtain that:

\begin{eqnarray}\label{Ql}
Q \simeq {Q_l} = \frac {\omega {L^2}} {{\pi^2} D} 
= \frac {\alpha\omega {L^2}} {c {\xi_0}}  
\end{eqnarray}
where the subscript $l$ is for localized modes, $\alpha$ is a constant 
of the order of unity and  depends on the ratio of $D$ and $\xi_0$ according 
to the localization theory. For both of the models studied here, we find the 
$\alpha$ can be chosen to be equal to 0.7.  From this we can get 
that the critical length  
$L_c=\sqrt{{\frac{1}{\alpha}}{\xi_0}{\xi_1}}\simeq\sqrt{{\xi_0}{\xi_1}}$ 
which is consistent with the Letokhov theory. Eq.(\ref{Ql})
is a very interesting result for laser physics because it is obtained by 
the comparison of  
the Letokhov and Lamb theories, and it directly gives the 
relationship of the quality 
factor $Q$ of a random system with the no-gain localization length $\xi_0$ 
of the system.

\par
In  the weak disorder limit ${\xi_0} \gg 
{{L_c^0}^2}/{\xi_1}  $, $Q \rightarrow Q_p$ and $ L_c \rightarrow L_c^0 $.  
In strong disorder limit ${\xi_0} \ll {{L_c^0}^2}/{\xi_1}  $, $Q 
\rightarrow Q_l$ and $ L_c \rightarrow \sqrt{\xi_0\xi_1} $. For cases where 
 ${\xi_0}$ is 
comparable to $  {{L_c^0}^2}/{\xi_1}  $,  both the effects of periodic 
background and   randomness will be important to determine the quality 
factor of 
such a system. Considering the $Q$ as the photon-resistance in the system, 
and if we assume that both effects are $independent$ with each other, we 
have that the total quality factor of the system to be:

\begin{eqnarray}\label{Q}     
Q={Q_p}+{Q_l}={\frac {\omega}{c}} \left(\frac{{\xi_1}L}{L_c^0} + 
               \frac{\alpha L^2}{\xi_0} \right) 
\end{eqnarray}

From the critical condition Eq.(\ref{LambCritical}), we have that:

\begin{eqnarray}\label{Lc}
{L_c}=-\frac{\xi_0 \xi_1}{2{\alpha L_c^0}} +
\sqrt{ {\left(\frac{\xi_0 \xi_1}{2{\alpha L_c^0}}\right)}^2 +
\frac{\xi_0 \xi_1}{\alpha }} 
\end{eqnarray}

In Fig.~6a~and~6b, we compare the theoretical predictions given by  
Eq.(\ref{Lc}) and by Zhang 
 \cite{Zhang} with our numerically calculated results for the 
classical 
many-layered model and the electronic tight-binding model respectively. 
Our numerical results shown in Fig.~6a and 6b strongly 
support Eq.(\ref{Lc}) to be the correct expression of the critical 
length $L_c$ for 
both  the weak and the strong random limits. In some other cases, the 
deviation 
can be as large as fifteen percent which is still very good considering 
the many approximations that have been introduced in derivation of 
Eq.(\ref{Lc}). One explanation for this deviation is that the two 
effects that were added in Eq.(\ref{Q}) are not totally $independent$, 
because the correlated 
scattering of the periodic background will affect both $Q_p$ and $Q_l$.

\section{Probability Distribution of Reflection Coefficient}

\par

Pradhan and Kumar \cite{PR} first obtained the probability  
distribution of 
the reflection coefficient  for a long system with randomness and gain, 
which is given by:

\begin{eqnarray}
P(x)=P \left(\frac{R-1}{2q}\right)=\left(\frac{2q}{R-1}\right)}^2
{\exp\left({\frac{-2q}{R-1}}\right)                         \label{PR}
\end{eqnarray}
where $x=\frac{R-1}{2q}$ and  $q={\xi_0}/{\xi_1}$.  We numerically 
calculated  
the $P(x)$ (or $P(R)$) for both models in cases that $q$ changes 
drastically. Our numerical features of $P(x)$ ($P(R)$) give some 
interesting new results.
According to Eq.(\ref{PR}), the maximum
probability of $P(x)$ (or $P(R)$) should appear at 0.5 (or $ R=q+1 $), and 
the distribution has a long tail of large  $x$ (or $R$).

For the many-layered model, when the wavelength is  near the band center 
($\lambda=360nm$), we have that $L_c^0/\xi_1 \simeq 7$ for different 
gains from Eq.(\ref{LC0}). 
When $q$ at the range of 0.01 to 50, and so ${\xi_0} < {L_c^0}^2/{\xi_1} $, 
the   $P(x)$ ( $P(R)$) behaves as predicted by Eq.(\ref{PR}).
When $q$ is at the range of 50 to 500, the position of the maximum $P(x)$ 
($P(R)$) 
begins to shift   left away from the point 0.5 (or $q+1$ ).
When $q$ increase further, such as to become close to a thousand, the 
maximum of $P(R)$ shift to the value of $R_0$,
which is the saturated value of reflection for the periodic system, and 
$P(R)$ begin to change its shape 
into a delta function and the long tail disappears. This process is 
clearly shown in Fig.~7. It is reasonable 
to assume that when $q$ is very large, the system is similar to
a periodic system , so the $P(R)$ changes to a delta function which is the
distribution of the reflection coefficient of the periodic system.

\par
For the tight-binding model, when frequency is at band center ($E=0$), 
${L_c^0}/{\xi_1}$ is not a constant as $\eta$ changes. It has a range from 
 5 to 7. When $q$ at
the range of 0.01 to 20(${\xi_0} < {L_c^0}^2/{\xi_1} $), the $P(x)$ 
( or $P(R)$ ) behaves as predicted by Eq.(\ref{PR}). When 
$q$ is at the range of 20-400, then the position of the maximum  $P(x)$ ( 
$P(R)$ ) begins to 
shift left away from theoretical value 0.5 ( or $q+1$). When $q$ is 
larger than 400, $P(R)$ develops two peaks 
, one peak  evolves from the original peak, the other one  emerges
at $R_0$. When $q$ is even larger, such as thousands, then the original   
peak goes down and disappears, and the new peak become higher at $R_0$. At
the same time the long tail disappears, the $P(R)$ also changes to a delta
function at the position of $R_0$. All these changes are shown clearly in 
Fig.~8. In Ref.\cite{India}, they also got two peaks
for $P(R)$, but they did not explain that the new peak is due to the
periodic
background of the system and that the delta function is at the position of
$R_0$, the saturated value of periodic system. When q is very small, we 
obtain that  the $P(x)$ ( or $P(R)$) is almost same as the one predicted by 
 Eq.(\ref{PR}), quite different from results 
of \cite{India}.  We think that
this difference is due to the fact that they have not renormalized their 
numerical results. 

\par
In summary, from our numerical results, we get the general behavior of 
 $P(x)$ ( or $P(R)$ ) for
both models. When ${\xi_0} < {L_c^0}^2/{\xi_1} $, the $P(x)$ ( or $P(R)$ ) is
same as the theoretically predicted one by Eq.(\ref{PR}). When 
${\xi_0}$ is bigger 
than ${L_c^0}^2/{\xi_1} $, we must think about the effect of the periodic 
background and if ${\xi_0}$ is really very large, the periodic background 
will 
dominate the behavior of $P(R)$. We also find that  at the band edge 
wavelength for many-layered model 
($\lambda$=470nm) or at the band edge energy for the tight-binding model 
$(E=1.8)$, 
the results of $P(x)$ ( or $P(R)$ )  are always different from the 
predictions given by Eq(\ref{PR}). 
We know that at the band edge the effects of coherent scattering will be 
very strong and make the long paths of wave propagation more important . 
So we think it is this coherent scattering effect which make the $P(x)$ ( 
or $P(R)$ ) different from the 
prediction of Eq(\ref{PR}) which is obtained for homogeneous random systems.

\section{Conclusion}

In conclusion, we have studied the transmission and reflection 
coefficients   in periodic or 
periodically correlated random systems  with homogeneous gain. 
Theoretically, for periodic systems we predicted the behaviors of 
transmission and reflection coefficients, such as the slopes of long 
systems and of short systems and critical length by the transmission matrix 
method. For  random systems, first the Zhang's formula of the
localization length for long systems is checked. We find that only 
at the band 
edge and with very strong gain and strong disorder, there is 
obvious deviation from the theoretical prediction of the localization 
length with gain.  For short systems, our numerical results show that 
our generalization of the formula of absorbing system is correct for 
amplifying systems.  According to this generalization we can predict the 
behaviors of average 
 value of logarithm of transmission coefficient $<\ln{T}>$ from the 
value of 
$1/\xi'$, such as: if it is positive then the $<\ln{T}>$ will increase from 
origin at slope $1/\xi'$ and generate a peak at $Lc$ and then start to 
decrease at slope $-1/\xi$, if it is negative then the 
$<\ln{T}>$ will decrease monotonically   and has a turning 
point at $L_c$ with the slope change from -$1/|\xi'|$ to $-1/\xi$. 
\par
To explain the new behavior of the critical length $L_c$ which we got from 
our numerical results, we 
compare the Letokhov theory with the Lamb theory and give a general 
expression for the  critical length considering both the effects of 
localization and periodic background. With this comparison, we also 
construct the relation of the quality factor $Q$ of a random system with the 
localization length $\xi$.           

\par
We also study the probability distribution of the reflection coefficient 
$P(R)$ of random systems with gain. We find some new behaviors of $P(R)$ and 
give the 
criteria for the range of validity of the  different behaviors 
and explain it by the influence of the periodic background too. 
The study of wave propagation in amplifying random system is a new and 
challenging topic. There are still a lot things to be done.


\section{Acknowledgment}

Ames Laboratory is operated for the US Department of Energy by Iowa State 
University by Iowa State University under Contract No.W-7405-Eng-82. This 
work was supported by the Director for Energy Research office of Basic 
Energy Sciences and Advanced Energy Projects and by NATO Grant No.940647.

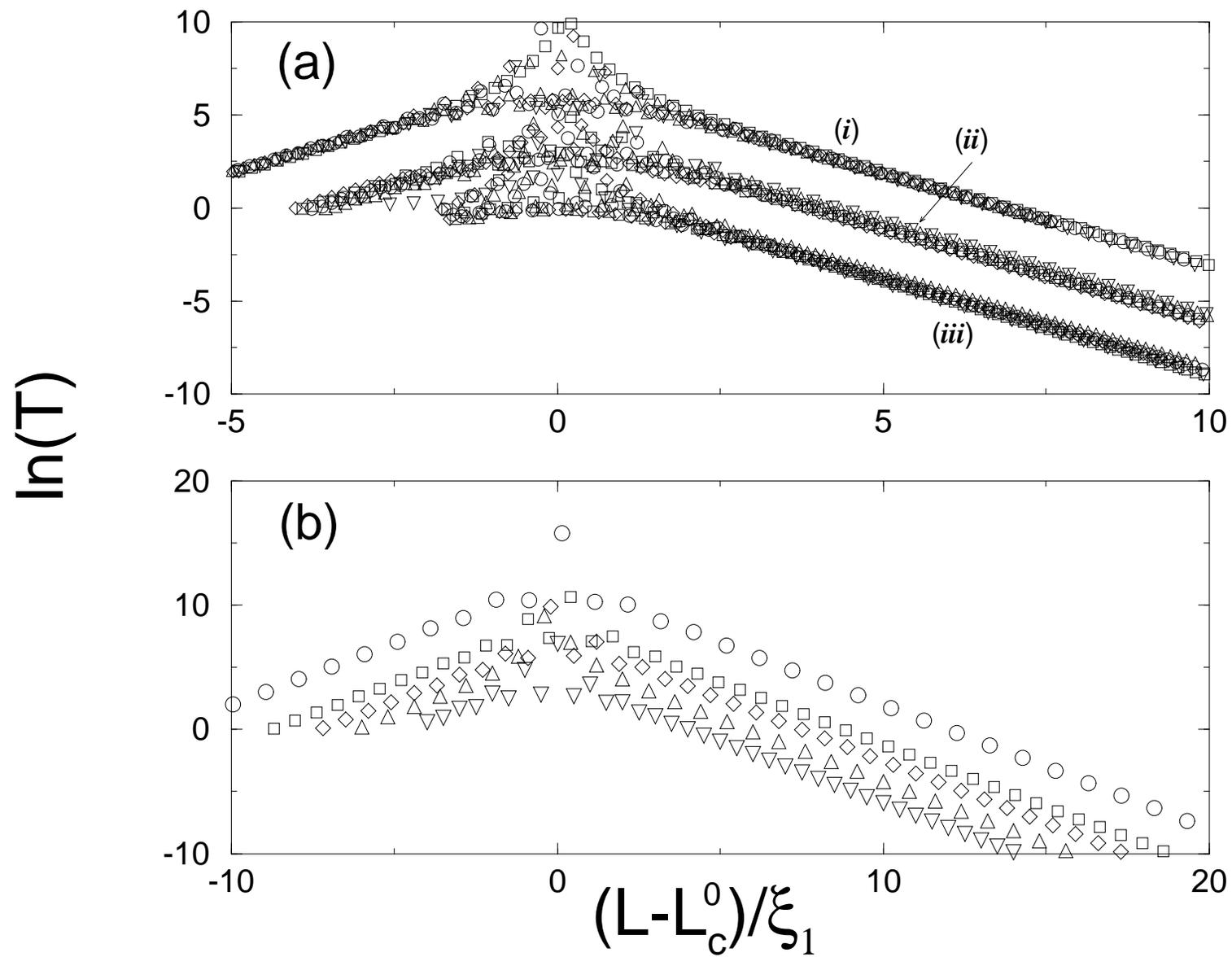
\begin{figure}[b]\label{T0}
\caption{The logarithm of the transmission coefficient $T$ versus
$(L-L_c^0)/\xi_1$, where $L_c^0$ is the
critical length and $\xi_1$ is the gain length of periodic
systems.
(a) For the periodic  many-layered model,  ($i$), ($ii$) and ($iii$) are 
the values at three representable  
wavelength   $\lambda$= 360 nm(band center), 420 nm(general) 
and 470 nm(band edge) respectively. The 
different symbols represent  values obtained from different gains,  
 $\varepsilon''$ = -0.001, -0.002, -0.005, -0.001, -0.1.
(b) For the periodic tight-binding model with E=0 for
different gains $\eta$=0.01, 0.05, 0.1, 0.2, 0.5. }
\end{figure}

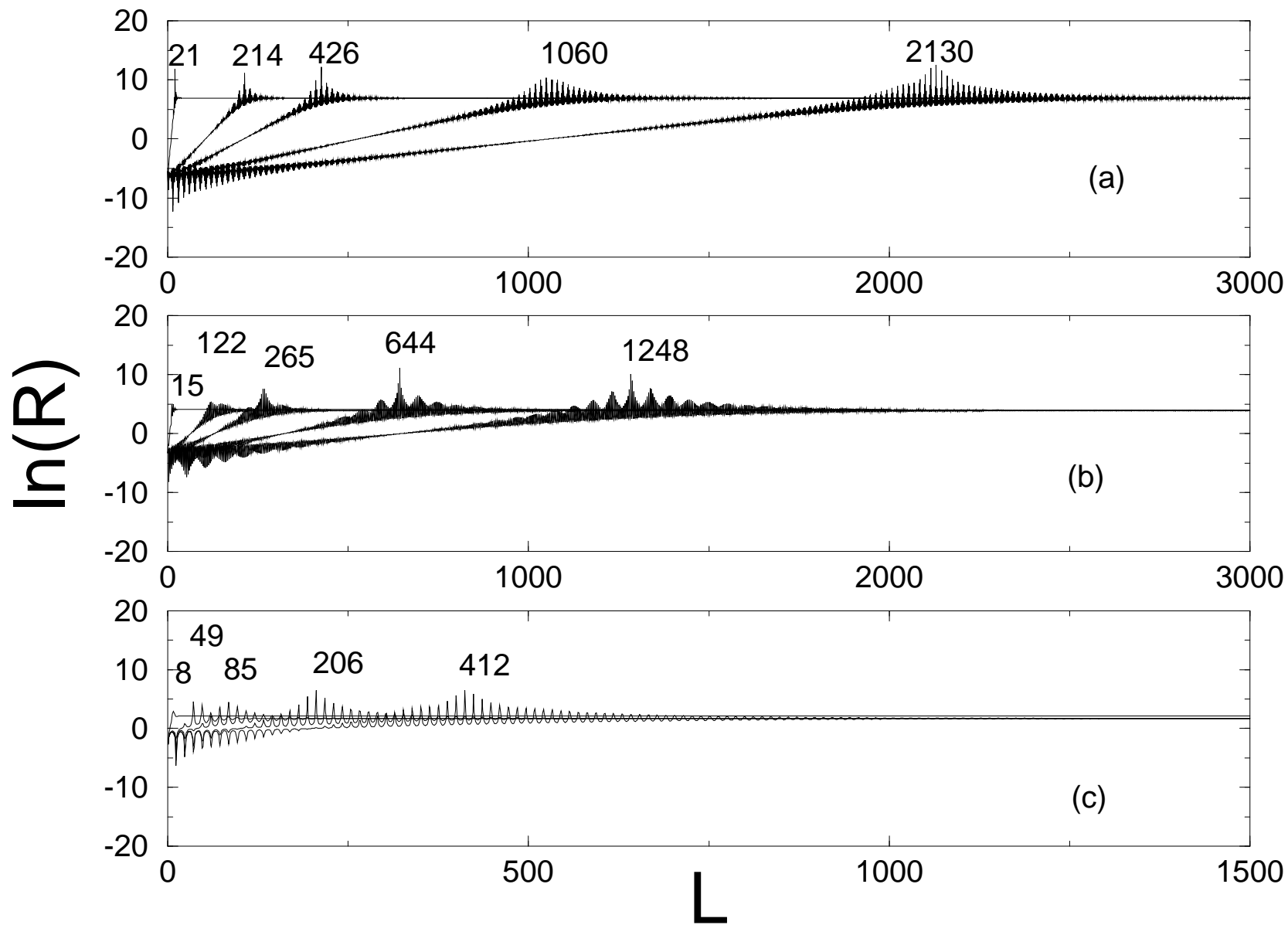
\begin{figure}[b]\label{R0}
\caption{The logarithm of the reflection coefficient $R$ versus L for 
the periodic many-layered model. (a), (b) and (c) are values of three 
representative wavelengths $\lambda$= 360 nm, 420 nm and 
470 nm respectively. From right to left, the numbers on the peaks are 
the values of  $L_c^0$,  
corresponding  to different gains $\varepsilon''$ = -0.001, -0.002, -0.005, 
-0.01, -0.1. Notice that the saturated value of R is independent of  
$\varepsilon''$ for the three wavelengths studied.} 
\end{figure}

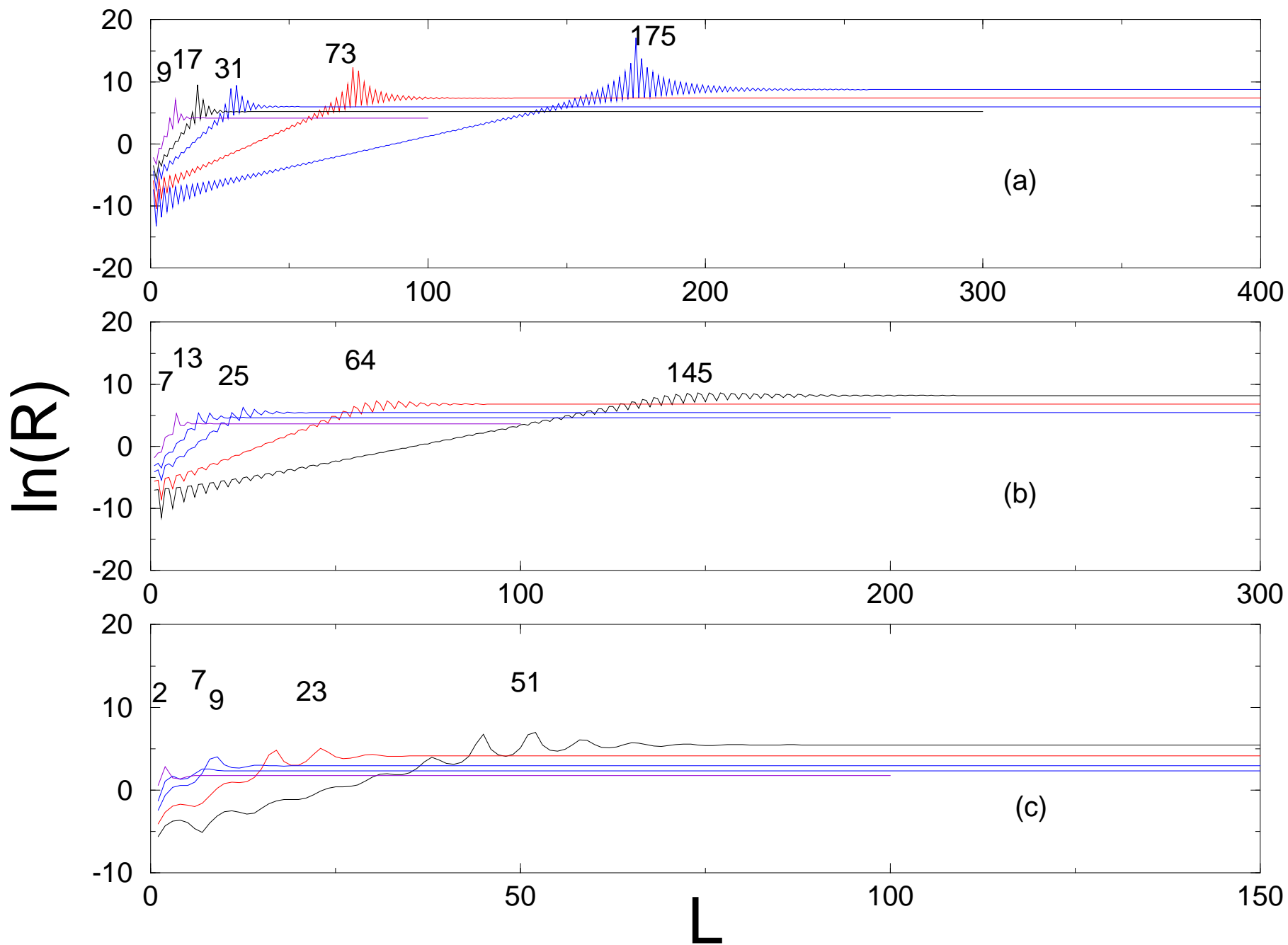
\begin{figure}[b]\label{R0tt}
\caption{The logarithm of the reflection coefficient $R$ versus 
L for periodic tight-binding 
model, (a), (b) and (c) are values of three representative energy E =0, 1, 
1.8. From right to left, the numbers  on the peaks are the values of  
$L_c^0$ corresponding  to different gains $\eta$ =0.01, 0.05, 0.1, 0.2, 
0.5. Notice that the saturated value of R for each E depends on $\eta$.}  
\end{figure}

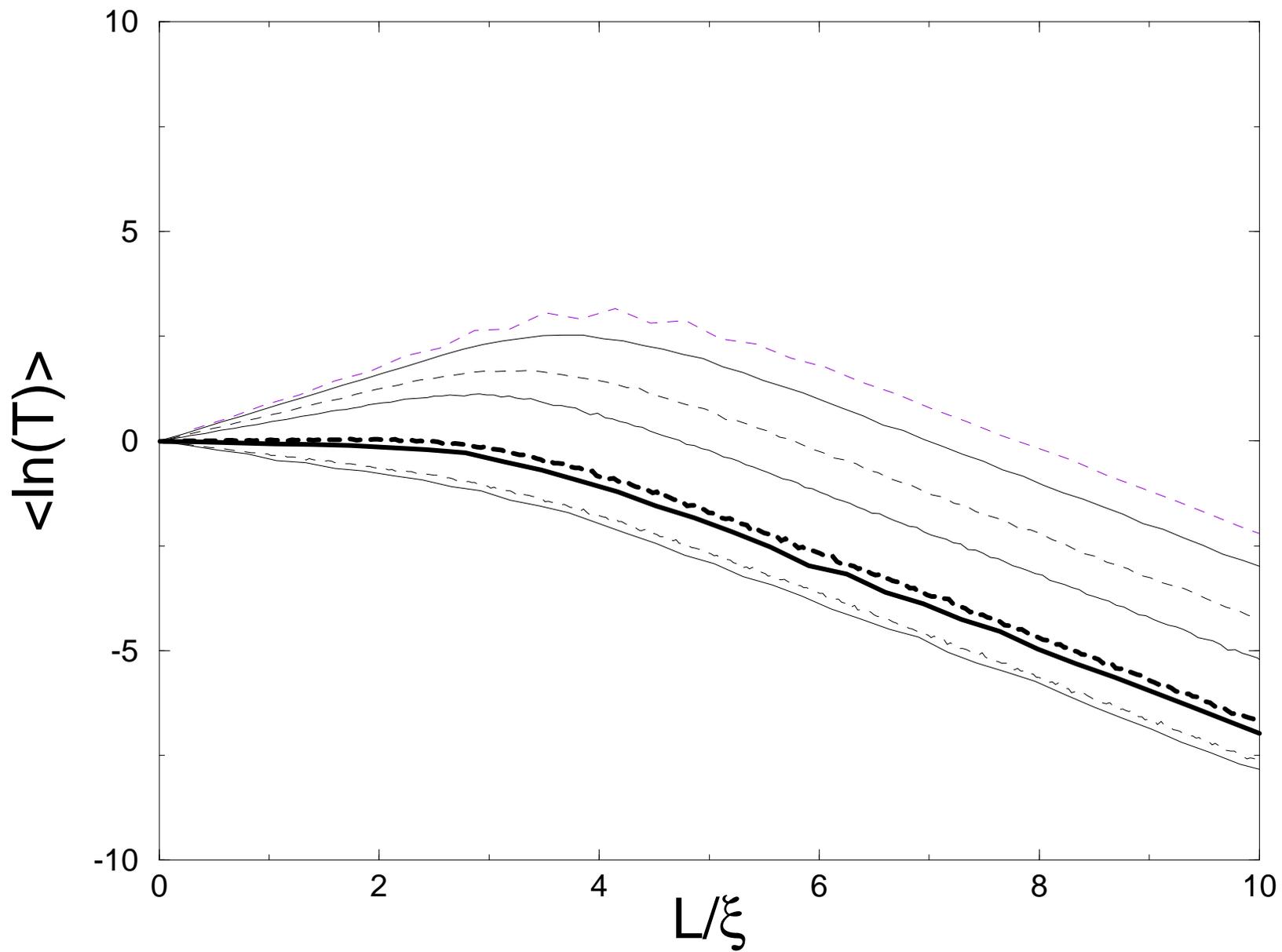
\begin{figure}[b]\label{gen}
\caption{The average values of logarithm
of $T$ versus $L/\xi$.
The results the random many-layered model, with $\lambda$=360 nm and W=0.2, 
are shown by
solid lines. Lines from lower to higher correspond to different gains 
$\varepsilon''$ = 0.0005, 0.001, 0.005, 0.01. When $\varepsilon''$ is 
equal to 0.001, $1/\xi_1$ is almost same as $1/\xi_0$, so $<\ln{T}>$  
is almost horizontal for small $L$, as shown by the wide solid line. For 
$\varepsilon''>0.001$, $1/\xi_1>1/\xi_0$, and for $\varepsilon''<0.001$, 
$1/\xi_1<1/\xi_0$. 
Results for the random tight-binding model,  with $E=0$ and $W=1$, are 
shown by dashed lines. Lines from lower to higher  correspond to  
$\eta$= 0.01, 0.02, 0.08, 0.3. When $\eta$ is equal to 0.02, $1/\xi_1$ is almost 
same as $1/\xi_0$,  so $<\ln{T}>$ is almost horizontal for small $L$, as shown 
by the wide dashed line. For $\eta>0.02$, $1/\xi_1>1/\xi_0$, and for 
$\eta<0.02$, $1/\xi_1<1/\xi_0$. } 
\end{figure}

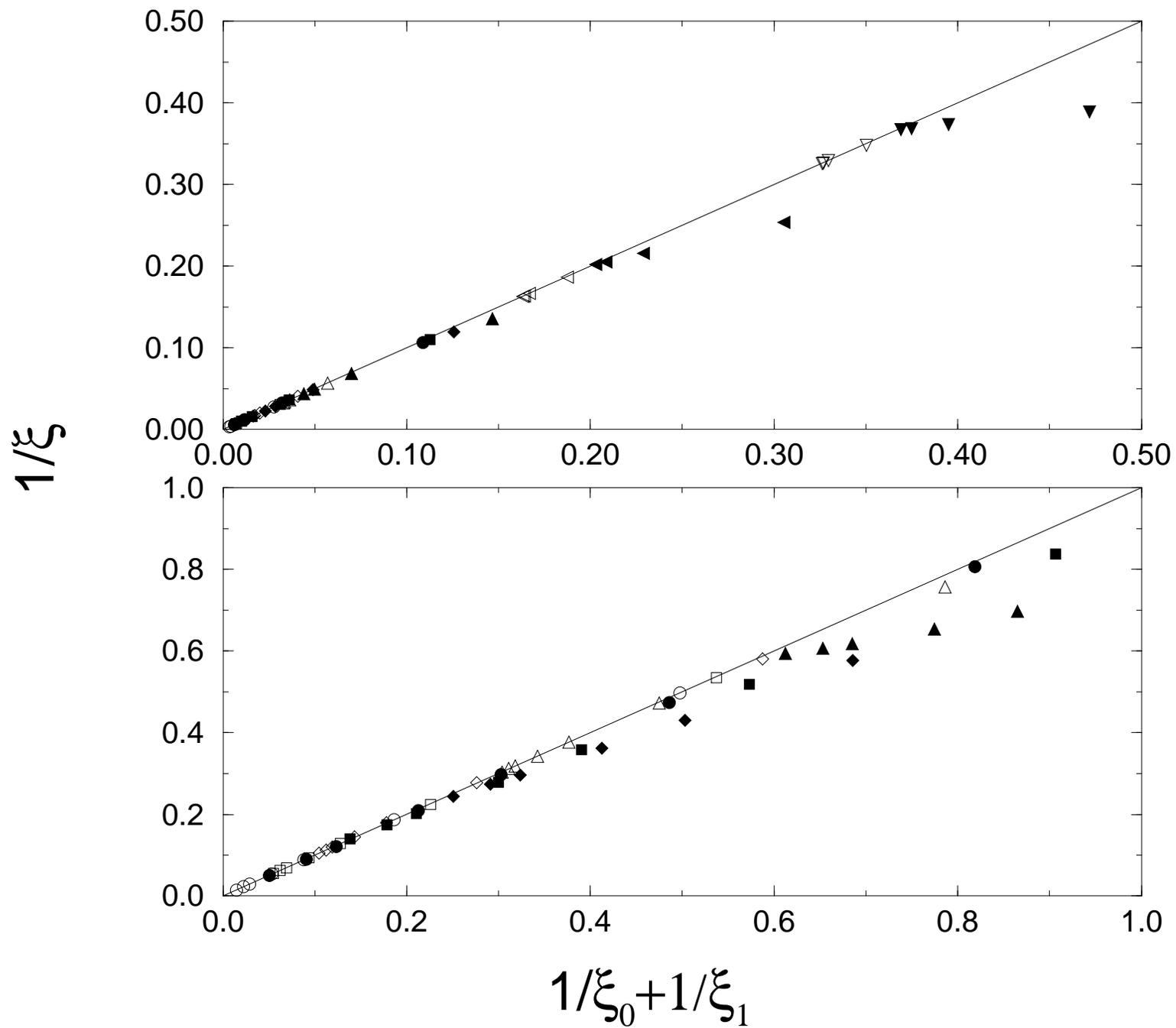
\begin{figure}[b]\label{figxi}
\caption{$1/\xi$
versus to $1/\xi_0 + 1/\xi_1$, where $\xi$ is localization length for a
system with gain, $\xi_0$ is the localization length of a  system
with disorder but with zero gain, and $\xi_1$ is the gain length.
(a) For the random many-layered model, empty symbols are of wavelength 
$\lambda$=360 nm, filled symbols are of $\lambda$=470 nm. Different symbols  
represent different sets of parameters of disorder and gain, $W=0.05, 
0.1, 0.2, 0.5$; $\varepsilon''$=0.001, 0.002, 0.005, 0.01, 0.05, 0.1.
(b) For the random tight-binding model,   empty symbols are of energy 
$E=0$, filled  symbols are of energy $E=1.8$. Different symbols  represent 
different sets of parameters of disorder and gain, 
$W$=0.5, 0.8, 1, 1.5, 2, 3, 5; $\eta$=0.01, 0.05, 0.1, 0.3. } 
\end{figure}

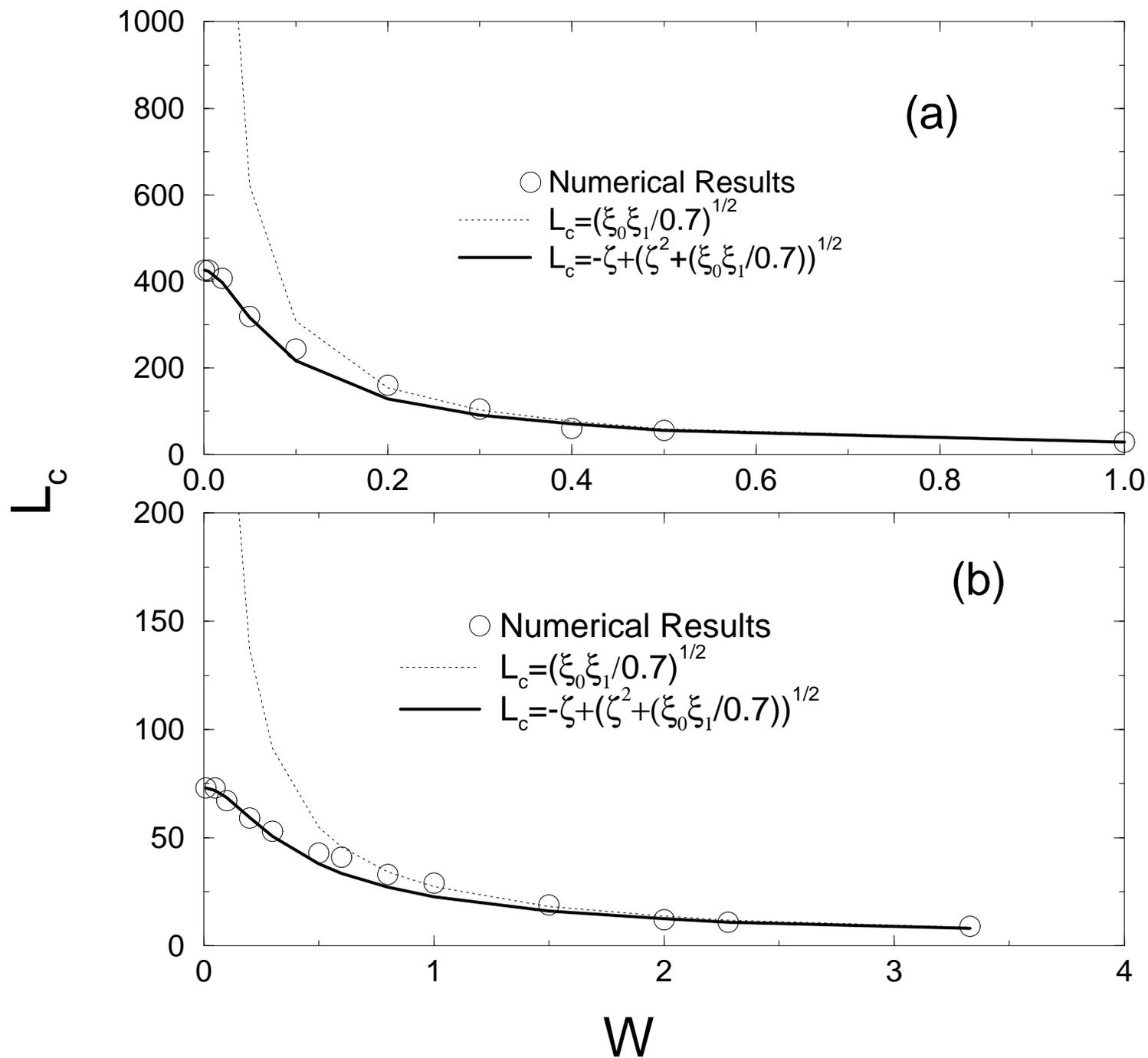
\begin{figure}[b] 
\caption{The critical length $L_c$ is plotted versus different
random strengths  $W$.
(a) For the random many-layered model with $\lambda$=360nm,  the 
dashed line and darkened line are the values obtained  according to Zhang's 
formula and Eq.(\ref{Lc})  respectively.  
(b) For the random tight-binding model with $E$=0 and $\eta$=0.01, 
the dashed line and darkened line are the values obtained  according to 
Zhang's formula and Eq.(\ref{Lc})  respectively. 
In both cases $\zeta=\frac{\xi_0\xi_1} {2 \alpha {L_c^0} }$ and 
$\alpha=0.7$.    } 
\end{figure}

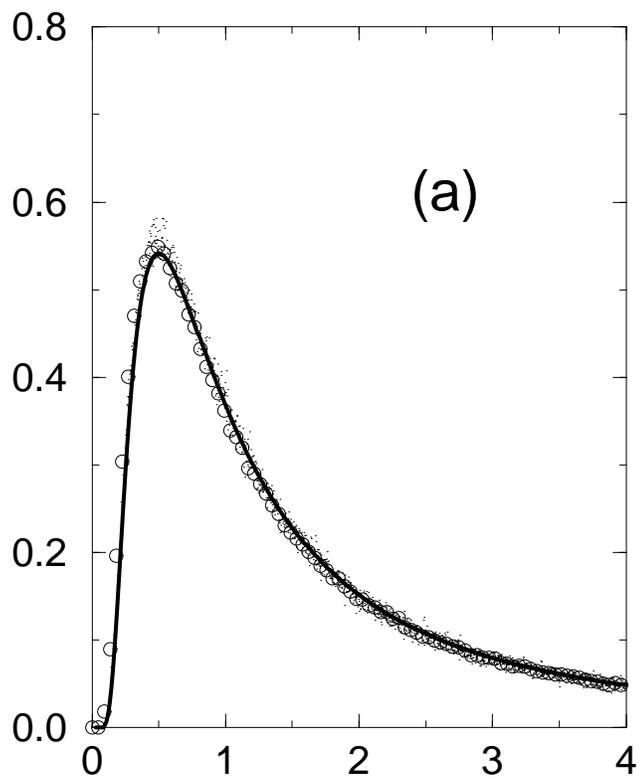
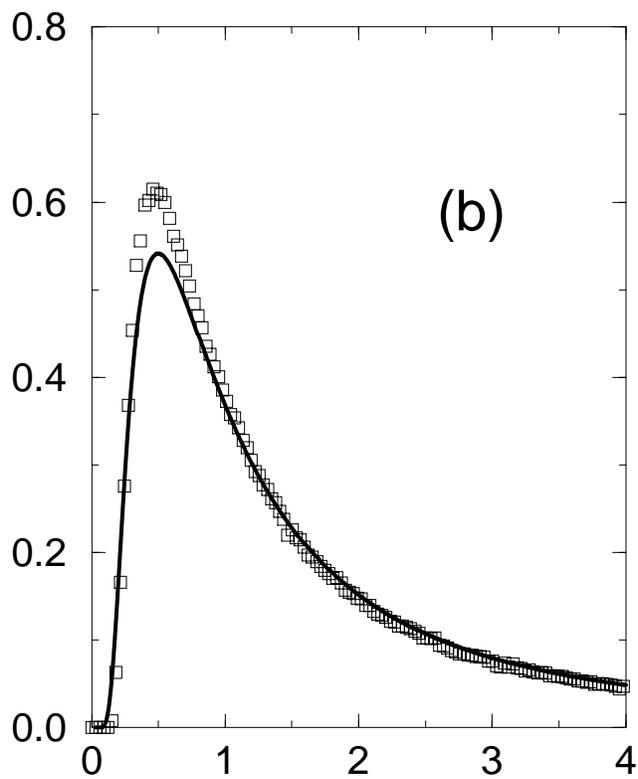
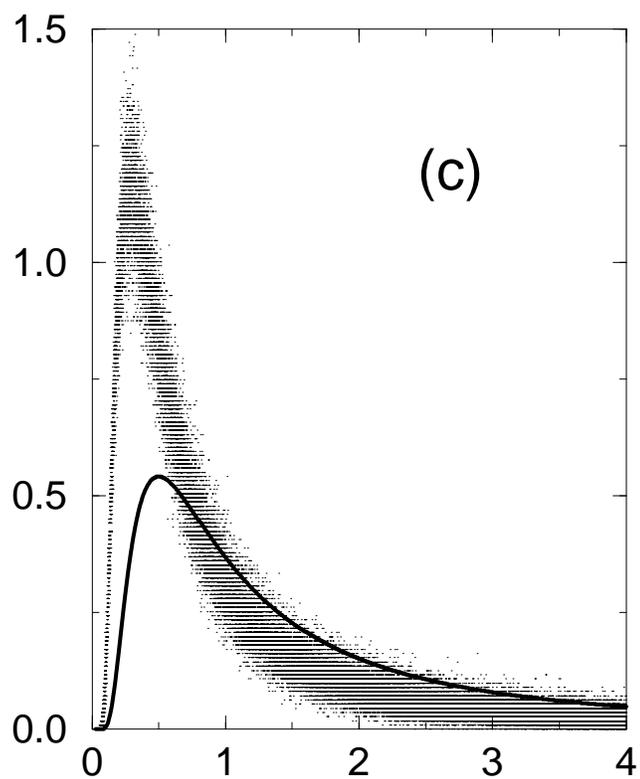
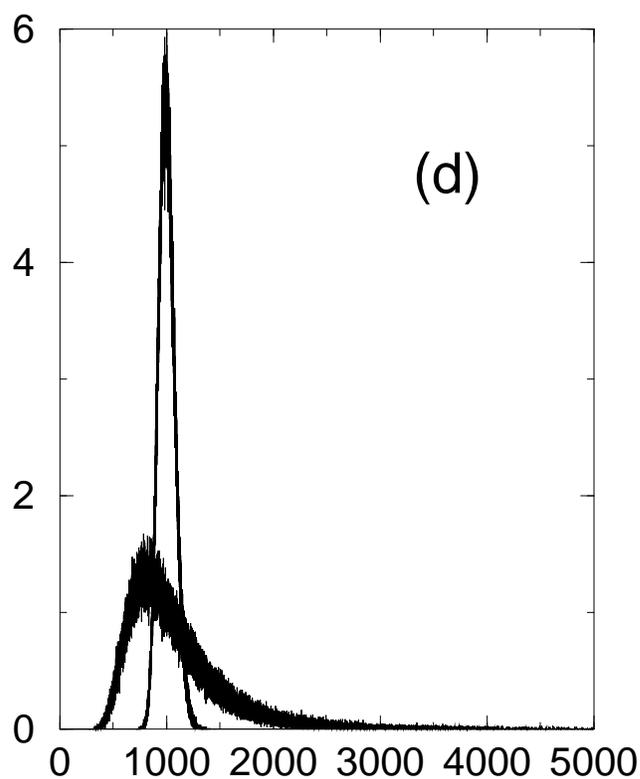
\begin{figure}[b] 
\caption{Probability distribution of the reflection coefficient P(x) versus x 
of the random many-layered system with gain at $\lambda=300$ nm for 
$q=1.1$ and $17.7$ (a); $q=0.163$ (b); $q=451$ (c),  where $x=(R-1)/q$ and 
$q=\xi_0/\xi_1$. The solid curve given by the solid line in (a), (b) and 
(c) is the analytical result of Eq.(26). In (d), $P(R)$ versus $R$ is 
plotted for two values of $q$ , $q=1800$(low one) and $7200$(high one). 
Notice that $P(R)$ approaches a delta-function distribution at $R_0$ when 
$q=7200$.} 
\end{figure} 

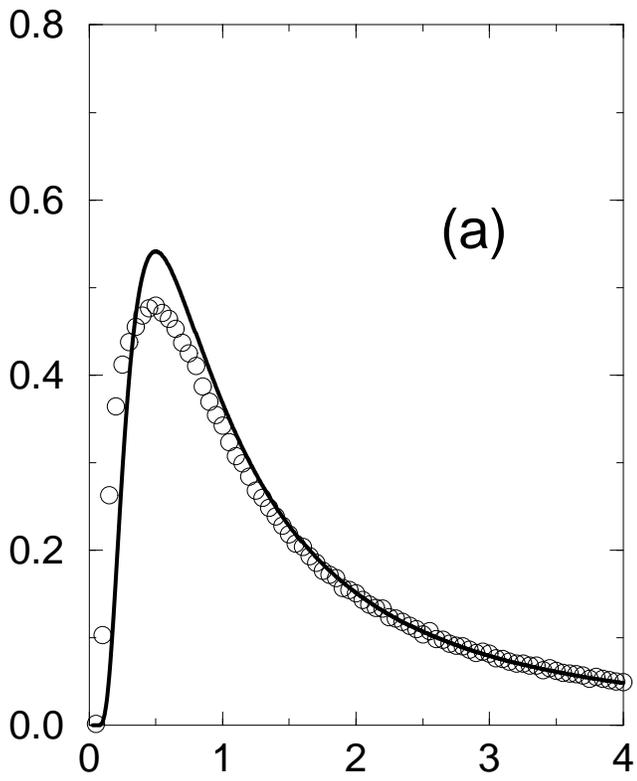
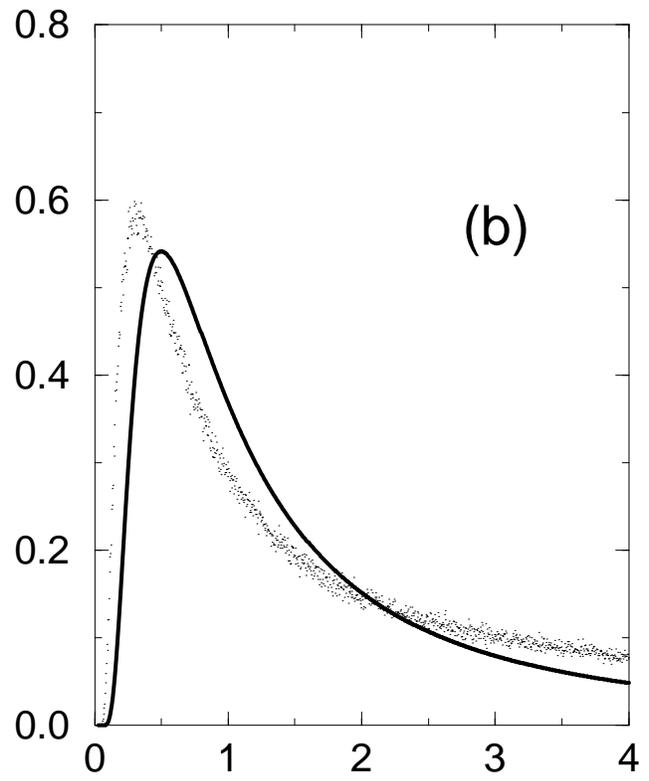
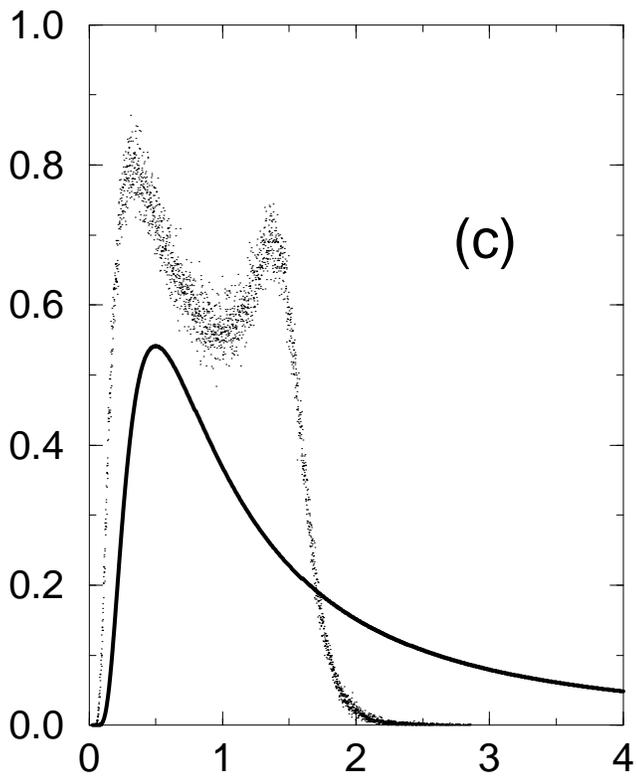
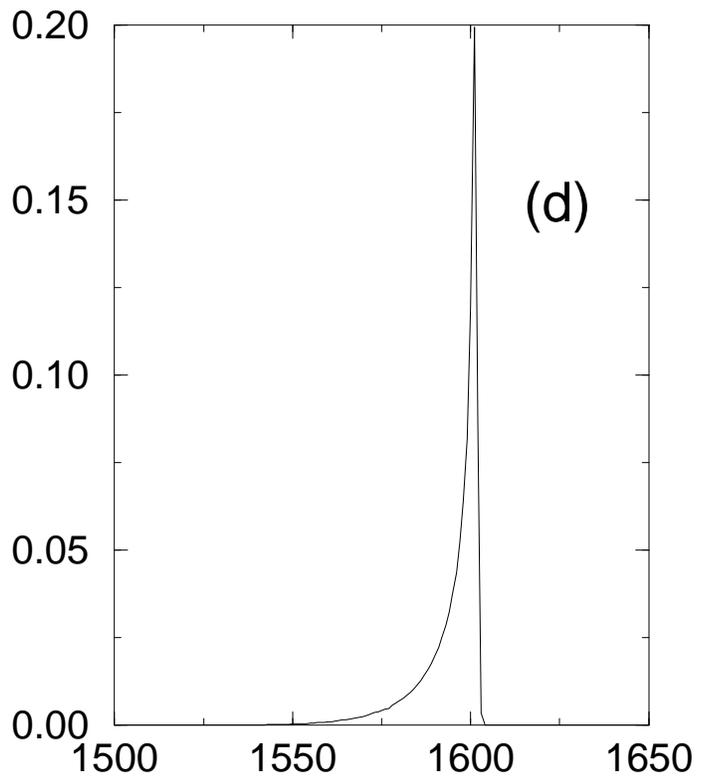
\begin{figure}[b] 
\caption{Probability distribution of the reflection coefficient P(x) versus x
of the random tight-binding system with gain at $E=0$ for $q=1.0$ (a); 
$q=132$ (b); $q=525$ (c), where $x=(R-1)/q$ and $q=\xi_0/\xi_1$. The 
solid curve given by the solid line in (a), (b) and (c) is the analytical 
result of Eq.(26). In (d), $P(R)$ versus $R$ is plotted for 
$q=5.25 \times 10^4$. $P(R)$ approaches a delta-function distribution 
at $R_0$ when $q=5.25 \times 10^4$.   } 
\end{figure}

\end{document}